\documentclass[a4paper,superscriptaddress,aps,reprint]{revtex4-2}
\usepackage[utf8]{inputenc}
\usepackage{lipsum}
\usepackage{xcolor}
\usepackage{amssymb,amsmath,amsthm,amsfonts,mathrsfs}
\usepackage{graphicx}
\usepackage{tabularx}
\usepackage{booktabs}
\usepackage[hidelinks]{hyperref}
\usepackage[capitalise]{cleveref}
\usepackage{xcolor}
\usepackage{subfig}
\usepackage{soul}
\usepackage{ragged2e} 

\usepackage{bbold}
\usepackage{comment}
\usepackage{csquotes}
\usepackage[normalem]{ulem}

\newtheorem{theorem}{Theorem}

\usepackage{physics}
\usepackage{braket}
\usepackage{caption}
\renewcommand\bra[1]{{\langle{#1}|}}
\makeatletter
\renewcommand\ket[1]{%
 \@ifnextchar\bra{\k@t{#1}\!}{\k@t{#1}}%
}
\newcommand\k@t[1]{{|{#1}\rangle}}
\newcommand{\be}{\begin{eqnarray}}
	\newcommand{\ee}{\end{eqnarray}}

\newcommand{\ch}{\mathcal{H}}
\def\phys{{\mathrm{phys}}}
\def\kin{{\mathrm{kin}}}
\usepackage{hyperref,color}

\makeatother
\newcommand{\new}[1]{{#1}}

\definecolor{darkblue}{RGB}{0,0,149}
\hypersetup{colorlinks=true,linkcolor=darkblue,filecolor=darkblue,urlcolor=darkblue,citecolor=darkblue,linktocpage=true}
\usepackage{url}

\begin{document}

\title{The sum of entanglement and subsystem coherence is invariant under quantum reference frame transformations}

\author{Carlo Cepollaro}
\affiliation{%
	Vienna Center for Quantum Science and Technology (VCQ), Faculty of Physics, University of Vienna, Boltzmanngasse 5, A-1090 Vienna, Austria
}%
\affiliation{%
	Institute of Quantum Optics and Quantum Information (IQOQI), Austrian Academy of Sciences, Boltzmanngasse 3, A-1090 Vienna, Austria
}%
\email{carlo.cepollaro@oeaw.ac.at}

\author{Ali Akil}
%\affiliation{%
%Department of Physics, Southern University of Science and Technology, Shenzhen 518055, China
%}%
\affiliation{%
Department of Computer Science, The University of Hong Kong, Hong Kong Island, Hong Kong S.A.R. China
}%
%\affiliation{%
%Jockey Club Institute for Advanced Study, The Hong Kong University of Science and Technology, Clear Water Bay, Kowloon, Hong Kong, China
%}%
\email{aliakil@hku.hk}

\author{Pawe{\l} Cie\'sli\'nski}
\affiliation{%
	 Institute of Theoretical Physics and Astrophysics, Faculty of Mathematics, Physics and Informatics, University of Gdańsk, 80-308 Gdańsk, Poland
}%

\author{Anne-Catherine de la Hamette}
\affiliation{%
	Vienna Center for Quantum Science and Technology (VCQ), Faculty of Physics, University of Vienna, Boltzmanngasse 5, A-1090 Vienna, Austria
}%
\affiliation{%
	Institute of Quantum Optics and Quantum Information (IQOQI), Austrian Academy of Sciences, Boltzmanngasse 3, A-1090 Vienna, Austria
}%

\author{\v{C}aslav Brukner}
\affiliation{%
	Vienna Center for Quantum Science and Technology (VCQ), Faculty of Physics, University of Vienna, Boltzmanngasse 5, A-1090 Vienna, Austria
}%
\affiliation{%
	Institute of Quantum Optics and Quantum Information (IQOQI), Austrian Academy of Sciences, Boltzmanngasse 3, A-1090 Vienna, Austria
}%
\email{caslav.brukner@univie.ac.at}

\begin{abstract}
Recent work on quantum reference frames (QRFs) has demonstrated that superposition and entanglement are properties that change under QRF transformations. Given their utility in quantum information processing, it is important to understand how a mere change of perspective can produce or reduce these resources. Here we prove the existence of a QRF invariant which can be decomposed such that it captures the trade-off between entanglement and subsystem coherence: We demonstrate the invariance of the sum of entanglement and subsystem coherence for two pairs of resource quantifiers. Moreover, we find a weaker trade-off that holds for any possible pair of measures. Finally, we discuss the implications of this interplay for violations of Bell's inequalities, clarifying that for any choice of QRF, there is a quantum resource responsible for the violation. These findings contribute to a better understanding of the quantum information theoretic aspects of QRFs, offering a foundation for future exploration in both quantum theory and quantum gravity. 
\end{abstract}

\maketitle

\section{Introduction}
The principle of covariance, which states that physical laws remain unchanged under a reference frame transformation, forms our fundamental understanding of the physical world. However, a variety of measurable physical quantities, such as energy, or magnetic and electric fields, are frame-dependent. Similar considerations arise when investigating the behavior of quantum systems relative to a \emph{quantum reference frame} (QRF). For example, a system that is well localized in space in one QRF may be in a superposition of different locations in a different QRF~\cite{Giacomini_2019}.
The same applies to quantum entanglement, making it a frame-dependent notion in quantum theory. Since both coherence and entanglement are recognized as useful resources in quantum information science, the question naturally arises: \textit{How can a mere change of perspective from one frame of reference to another create a useful resource?} 

In this paper we give a natural answer to this question by showing that, under QRF transformations, there exists an invariant which can be decomposed such that it captures a trade-off between the entanglement and subsystem coherence, in the basis associated with the symmetry group of the transformation: 
We show that, for certain entanglement and coherence quantifiers, their sum is preserved under QRF changes, for any pure bipartite state. \new{Thus, any entanglement increase under a QRF transformation comes at the expense of a decrease in coherence, and vice versa.} Furthermore, we showed that this property does not hold for all quantifier pairs: fixing an entanglement (or coherence) quantifier, there is at most one coherence (or entanglement) quantifier that yields an invariant sum.

The apparent \enquote{creation} of coherence or entanglement under a QRF change can then be explained by there existing either one of these resources in the original frame, and the change of frame transforming one into the other. This is analogous to the case of electrodynamics, where a magnetic field can appear when applying a Lorentz boost, as a consequence of an electric field being present in the rest frame. As Maxwell showed, they are two sides of the same coin, just described in different frames: they are different components of the electromagnetic tensor. Here we show that coherence and entanglement exhibit a similar behavior under QRF transformations. Moreover, just like in electromagnetism, where there is a quantity combining both magnetic and electric fields that is left invariant under Lorentz boosts (the free-field Lagrangian density), here we identify meaningful quantities that are left invariant under QRF changes. 

To demonstrate the significance of the trade-off between coherence and entanglement in the field of quantum information, we discuss violations of Bell's inequalities and clarify the potential misunderstandings regarding \enquote{production} or \enquote{destruction} of resources through QRF transformations. Different QRFs induce different factorizations of the Hilbert space, both for the states and for the observables \cite{Ahmad2022, delahamette2021perspectiveneutral, castroruiz2021relative, hoehn2023quantum, devuyst2024gravitationalentropyobserverdependent}, such that, for example, entangled states measured by local operations (as in the Bell experiment) transform into product states and non-separable operations. Thus the resource of entanglement is transferred from the state to the observables.

The study of various notions of reference frames in quantum theory has been pursued through different approaches within the contexts of superselection rules 
and quantum information protocols~\cite{Kitaev2004, Bartlett2007, Gour_2008, Palmer14}, quantum gravity~\cite{Rovelli_1991, Dittrich_2006,Hohn2019}, and symmetries~\cite{loveridge_relativity_2017, loveridge_symmetry_2018}. During a more recent wave of research, several approaches to QRFs have been formulated~\cite{Giacomini_2019, delaHamette2020quantumreference, Vanrietvelde2020, hoehn2019trinity, castroruiz2021relative, delahamette2021perspectiveneutral, carette2023operational, hoehn2023quantum} and applied at the interface of quantum physics and gravity~\cite{giacomini2021einsteins, Giacomini_2021, delaHamette2021falling, Kabel2022conformal, apadula2022quantum, kabel2023quantum,cepollaro2024quantum,kabel2024identification, fewster2024quantumreferenceframesmeasurement, devuyst2024gravitationalentropyobserverdependent}.

One particular question we address in this paper is relevant to most of these approaches: \textit{If both superposition and entanglement are each frame-dependent notions on their own, is there any physical quantity involving the two that is invariant under quantum frame changes?} A first step to answering this question was taken by Savi and Angelo~\cite{Savi2021}, who found that the so-called \enquote{total quantumness} of the system is conserved under QRF changes. However, the conservation of \enquote{total quantumness} was shown using only the unitarity of QRF transformations, and not their specific form - the same result thus holds for any possible unitary transformation. In the present work, we show that the specific form of (ideal) QRF transformations~\cite{Giacomini_2019, delaHamette2020quantumreference} leaves the sum of entanglement and subsystem coherence invariant. \new{Despite different motivations, at the technical level our result is closely related to quantum resource theory findings in Refs.~\cite{Strelstov2015, Zhu_2017}. These results can be recovered by restricting to finite cyclic QRF transformations and vanishing initial entanglement, suggesting that the coherence–entanglement link identified in Refs.~\cite{Strelstov2015, Zhu_2017} admits an interpretation in terms of invariance under QRF transformations.} \new{See also Refs~\cite{Tessier_2005,Wang:2010qq} for related studies on trade-offs between single-system and bipartite properties in multi-qubits systems.}

The paper is structured as follows: Sec.~\ref{sec:example} introduces the essential features of QRFs with a simple qubit example. Sec.~\ref{sec:main} presents the general definition of QRF transformations, and states our main results: a conservation theorem for two quantifier pairs, along with a weaker trade-off for any pair. Proof ideas are outlined, with details in the Supplementary Material. Sec.~\ref{sec:protocols} discusses Bell inequality violations in QRFs, and Sec.~\ref{sec:conclusions} concludes with open questions.

\section{A simple example}
\label{sec:example}
Let us start by a simple example to build some intuition for changes of classical and quantum reference frame (cf.~\cite[Ex.~1]{delaHamette2020quantumreference}). Consider three systems $A$, $B$, and $C$, which can, for now, be in two states, $\uparrow$ or $\downarrow$. We can conventionally choose that each system considers themselves to be in the state $\uparrow$, similarly to how an astronaut freely floating in space would consider the direction \enquote{up} to be aligned from their toes to their head. Take the classical systems to be in the state $\uparrow^{(A)}_A \uparrow^{(A)}_B \downarrow^{(A)}_C$ relative to $A$. Then, the corresponding state relative to $B$ is $\uparrow^{(B)}_B \uparrow^{(B)}_A \downarrow^{(B)}_C$, while that relative to $C$ is $\uparrow^{(C)}_C \downarrow^{(C)}_A \downarrow^{(C)}_B$. Note how the \emph{relative} configurations between the systems are left invariant under these classical changes of frame.

Let us now extend this to the case of \emph{quantum} reference frames (cf.~\cite[Ex.~2]{delaHamette2020quantumreference}). For this, take $A$, $B$, and $C$ to be quantum systems with Hilbert space $\mathbb{C}^2$, and basis states $\ket{\uparrow}$ and $\ket{\downarrow}$. Correspondingly, if the state relative to $A$ is $\ket{\psi}^{(A)}_{ABC}=\ket{\uparrow}^{(A)}_A \ket{\uparrow}^{(A)}_B\ket{\downarrow}^{(A)}_C$, then the state relative to $B$ is $\ket{\psi}^{(B)}_{ABC}=\ket{\uparrow}^{(B)}_B \ket{\uparrow}^{(B)}_A \ket{\downarrow}^{(B)}_C$ and that relative to $C$ is $\ket{\psi}^{(C)}_{ABC}=\ket{\uparrow}^{(C)}_C \ket{\downarrow}^{(C)}_A \ket{\downarrow}^{(C)}_B$.
Thus, $A$, $B$, and $C$ can be understood as qubits described \emph{relative} to one another, meaning that $\ket{\psi}^{(C)}_{ABC}$ encodes the relative degrees of freedom of $A$ and $B$ with respect to $C$~\cite{Giacomini_2019}.

\begin{figure}[ht]
 \includegraphics[scale=0.7]{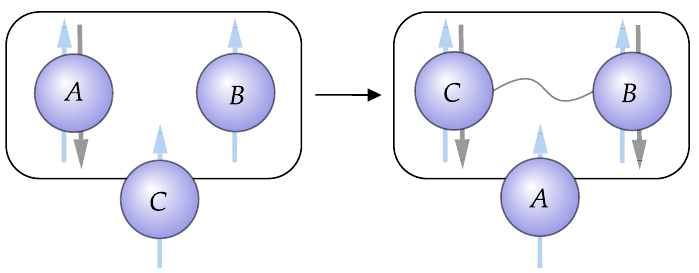} 
 \caption{\justifying Three qubits $A$, $B$, and $C$ are described relative to one another. On the left, systems $A$ and $B$ are described relative to system $C$. While $C$ is in the state $\protect\ket{\uparrow}$ relative to itself, $A$ is in a coherent superposition of $\protect\ket{\uparrow}$ and $\protect\ket{\downarrow}$ and $B$ is in the state $\protect\ket{\uparrow}$. Coherently changing reference frame to that of $A$, we find that $B$ and $C$ are in an entangled state. In general, superposition and entanglement depend on the QRF.}
 \label{fig:simple_example}
\end{figure}

Take now the state
\begin{equation}\label{eq:example_C}
\ket{\psi}^{(C)}_{ABC}=\ket{\uparrow}^{(C)}_C \frac{\ket{\uparrow}^{(C)}_A+\ket{\downarrow}^{(C)}_A}{\sqrt{2}}\ket{\uparrow}^{(C)}_B.
\end{equation} 
We can infer the corresponding state $\ket{\psi}^{(A)}_{ABC}$ by applying the principle of \emph{coherent change of reference frame}, namely in each branch we apply the classical reference frame transformation described above. The result is that the state relative to $A$ is
\begin{equation}\label{eq:example_A}
\ket{\psi}^{(A)}_{ABC}=\ket{\uparrow}^{(A)}_A \frac{\ket{\uparrow}^{(A)}_B \ket{\uparrow}^{(A)}_C+\ket{\downarrow}^{(A)}_B\ket{\downarrow}^{(A)}_C}{\sqrt{2}}.
\end{equation}
The transformation between Eq.~\eqref{eq:example_C} and Eq.~\eqref{eq:example_A} is a QRF transformation and is illustrated in Fig.~\ref{fig:simple_example}.

Note, furthermore, that the conventional state of the reference frame with respect to itself can be omitted, as it does not contain any information, being always $\ket{\uparrow}$. We can then represent states of two systems relative to the third in the Hilbert space $\mathcal{H}=\mathbb{C}^2 \otimes \mathbb{C}^2$. Furthermore, for future convenience, we identify the basis $\{\ket{\uparrow}, \ket{\downarrow} \}$ with the computational basis $\{\ket{0}, \ket{1}\}$, since they are isomorphic. In this simplified and more abstract notation, we can write 
\begin{equation}\label{eq:example_C_qubit}
\ket{\psi}_{AB}^{(C)} = \frac{\ket{0}_A+\ket{1}_A}{\sqrt{2}}\ket{0}_B
\end{equation}
and correspondingly
\begin{equation}\label{eq:example_A_qubit}
 \ket{\psi}^{(A)}_{BC} = \frac{\ket{0}_B\ket{0}_C+\ket{1}_B\ket{1}_C}{\sqrt{2}}.
\end{equation}
The transformation between the two states can be implemented with the unitary
\begin{equation}
 S^{(C)\to (A)} = \ket{0}_{CA}\bra{0} \otimes \mathbb{1}_B + \ket{1}_{CA}\bra{1} \otimes \sigma^x_B,
\end{equation}
meaning that the QRF transformation for these systems amounts to a quantum controlled flip, or CNOT, and can thus be seen as a refactorization of the Hilbert space. Note, however, that while it maps to a specific perspective relative to a single qubit, all measurements and operations are still performed in the laboratory frame.

While the state of $A$ relative to $C$ in Eq.~\eqref{eq:example_C_qubit} has some coherence (in the computational basis), the total state is factorized, that is, there is no entanglement between $A$ and $B$ relative to $C$. On the contrary, the state in the perspective of $A$ in Eq.~\eqref{eq:example_A_qubit} is maximally entangled, and as such each subsystem is maximally mixed, meaning it has no coherence. Thus, the QRF transformation maps a factorized state with subsystem coherence to an entangled state with zero subsystem coherence.
This is an instance of a generic feature of QRF transformations, which we prove as our main result: We show that the sum of entanglement and subsystem coherence is invariant under QRF transformations, for specific choices of entanglement and coherence quantifiers. Additionally, if there exists a frame in which either the coherence or the entanglement vanishes, a trade-off holds for any possible pair of measures.

\section{Main results}\label{sec:main}
Let us consider the most general ideal QRF transformation: Given a locally compact group $G$ and a Hilbert space $\mathcal{H} = L^2(G) \otimes L^2(G)$, the most general QRF transformation can be written as~\cite{delaHamette2020quantumreference} 
\be \label{eq:QRFT}
 S^{(C)\to (A)} = \int_G dg \, \ket{g^{-1}}_{CA}\bra{g} \otimes U_B^\dagger(g),
\ee
with $\{\ket{g}\}_{g \in G}$ being the basis of $L^2(G)$, i.e.~$\braket{g|g^\prime}=\delta(g^{-1}\circ g^{\prime})$, and $U(g)$ the left regular representation of $G$, such that $U(g)\ket{g^\prime} = \ket{g \circ g^\prime}$. In the case of discrete groups the integrals can be replaced by sums and the Hilbert space $L^2(G)$ with $\mathbb{C}[G]$. For instance, in the illustrative example above, the group is $G=\mathbb{Z}_2$ and $\mathcal{H}=\mathbb{C}[\mathbb{Z}_2] \cong \mathbb{C}^2$.

Given a generic bipartite pure state in the perspective of $C$, namely 
\be \label{eq:state_perspective_C}
\ket{\psi}^{(C)}_{AB} = \int_G dg \, dg^\prime\, \psi^{(C)}_{AB}(g,g^\prime) \ket{g}_A \ket{g^\prime}_B,
\ee 
we can calculate what the state in the perspective of $A$ is, namely
\begin{align}
 \ket{\psi}_{BC}^{(A)} &= S^{(C)\to (A)} \ket{\psi}_{AB}^{(C)} \nonumber \\
 &= \int_G dg \, dg^\prime\, \psi^{(A)}_{BC}(g,g^\prime) \ket{g}_A \ket{g^\prime}_B, \label{eq:state_perspective_A}
\end{align} 
with $\psi^{(A)}_{BC}(g,g^\prime) = \psi^{(C)}_{AB}(g^{-1},g^{-1} \circ g^\prime)$.

Upon direct application of the definition, QRF transformations can be shown to be incoherent operations (see Supplementary Material~A), i.e.~they cannot increase the coherence of a bipartite state - and in particular they preserve it. Nevertheless, when studying a subsystem only, QRF transformations can increase or decrease its coherence, as pointed out in the previous example. Hence, it is meaningful to characterize the amount of subsystem coherence before and after the QRF change, as well as the entanglement of the bipartite state, which can change as well. 

To begin, we study entanglement entropy and relative entropy of coherence as our resource measures. The entanglement entropy is defined as $\mathcal{ E }_e\!\left[\ket{\psi}_{AB}\right ] = \mathcal{S}[\rho_A]$, where $\rho_A = \text{Tr}_B[\ket{\psi}_{AB}\bra{\psi}]$ is the reduced density matrix of the subsystem $A$ and $\mathcal{S[\rho]} = -\text{Tr}[\rho \log \rho]$ is the von Neumann entropy. The relative entropy of coherence is $\mathcal{C}_e[\rho] = \mathcal{S}[\rho_d] - \mathcal{S}[\rho]$, where $\rho_d$ is the state obtained from $\rho$ by keeping only the diagonal terms in the group element basis $\{\ket{g}\}_{g \in G}$. Intuitively, $\mathcal{C}$ quantifies how large the off-diagonal elements of $\rho$ are in the given basis.

Given the entanglement in the perspective of $C$, i.e.~$\mathcal{E}_e^{(C)} = \mathcal{E}_e[\ket{\psi}_{AB}^{(C)}]$, the coherence of subsystem $A$ in the perspective of $C$, i.e.~$\mathcal{C}_e^{(C)}=\mathcal{C}_e[\rho_A^{(C)}]$, where $\rho_A^{(C)} = \text{Tr}_B[\rho_{AB}^{(C)}]$, and analogous definitions for the perspective of $A$, namely $\mathcal{E}_e^{(A)} = \mathcal{E}_e[\ket{\psi}_{BC}^{(A)}]$ and $\mathcal{C}_e^{(A)}=\mathcal{C}_e[\rho_C^{(A)}]$, the following holds:
\begin{equation} \label{eq:conservation1}
 \mathcal{C}_e^{(C)} + \mathcal{E}_e^{(C)} = \mathcal{C}_e^{(A)}+\mathcal{E}_e^{(A)}.
\end{equation}
The proof relies on showing that $\rho_{A}^{(C)}$ has the same diagonal elements as $\rho_{C}^{(A)}$ up to a permutation (see Supplementary Material~B), and hence the entropy of the diagonal states is the same before and after the transformation. 
Another consequence of the above is an additional conservation law, with a different pair of coherence and entanglement quantifiers: given the $l^2$-norm of coherence~\footnote[1]{Note that this function has been shown to violate monotonicity under incoherent operations and can thus not be considered as a valid measure. However, it still captures the coherence present in the state (see~\cite{Luders2021, Ma2021, Yu2019, Anand2021} for recent examples), similar to the correlation length (see e.g. \cite{correlation_length}) in entanglement detection.} $\mathcal{C}_{l^2}[\rho] = \sum_{g\neq h} \left |\rho_{g,h} \right |^2$  and the linear entropy of the subsystem $\mathcal{E}_l[\ket{\psi_{AB}}]=1-\Tr[{\rho_A}^2]$, it holds that
\be \label{eq:conservation2}
 \mathcal{C}_{l^2}^{(C)} + \mathcal{E}_l^{(C)} = \mathcal{C}_{l^2}^{(A)}+\mathcal{E}_l^{(A)}.
\ee
The proofs of Eqs.~\eqref{eq:conservation1} and \eqref{eq:conservation2} are detailed in Supplementary Materials~C1 and~C2. We formulate our main result in the form of a theorem:
\begin{theorem}\label{th:th1}
The sum of entanglement and subsystem coherence $\mathcal{E}+\mathcal{C}$ is invariant under QRF transformations of pure bipartite states, for $(\mathcal{E},\mathcal{C})=(\mathcal{E}_e,\mathcal{C}_e)$ or $(\mathcal{E}_l,\mathcal{C}_{l^2})$.
\end{theorem}

Our theorem demonstrates the invariance of the sum of entanglement and subsystem coherence under QRF changes, for any pure bipartite state.
Moreover, in Supplementary Material~C3, we prove that the pairs of quantifiers in Theorem 1 cannot be chosen arbitrarily: for each entanglement (or coherence) quantifier, there is at most one coherence (or entanglement) quantifier such that their sum is conserved. This provides insight into which quantifiers can form a QRF invariant.

Theorem 1 entails a trade-off between the two quantities: a QRF transformation that increases entanglement only does so by consuming the existing coherence, and vice versa. In other words, it follows from Eqs.~\eqref{eq:conservation1} and \eqref{eq:conservation2} that for both choices of quantifiers, when coherence increases under a change of frame, entanglement decreases, and vice versa, i.e. $\Delta \mathcal{C}_e \Delta \mathcal{E}_e\leq 0$ and $\Delta \mathcal{C}_{l^2} \Delta \mathcal{E}_{l}\leq0$, with $\Delta \mathcal{C}_e = \mathcal{C}_e^{(A)}-\mathcal{C}_e^{(C)}$, and similarly for the others. Note, however, that even this weaker trade-off does not hold for any pair of coherence and entanglement quantifiers
- a numerical example where $\Delta C_{l^2} \Delta \mathcal{E}_e\geq 0$ is given in Supplementary Material~C4. This means that there is a pair of measures such that coherence and entanglement can both increase under a QRF change. This is connected to the property of different entanglement and coherence measures not being monotones with respect to each other (see Refs.~\cite{VIRMANI200031,Liu2016}): they do not order states such that, given two different measures $\mathcal{C}_1$ and $\mathcal{C}_2$, for any two pairs of states such that $\Delta \mathcal{C}_1 <0$, it is also true that $\Delta \mathcal{C}_2<0$. 

Nevertheless, it is still possible to find a trade-off for any possible pair of measures, for some specific set of states. This can be formulated as follows.

\begin{theorem} \label{th:th2}
For any choice of entanglement measure $\mathcal{E}$ and coherence measure $\mathcal{C}$, if there is a QRF where a pure bipartite state has $\mathcal{E}=0$ or $\mathcal{C}=0$, under a QRF transformation it holds that $\Delta \mathcal{E} \Delta \mathcal{C}\leq0$.
\end{theorem}
The proof is analogous of the one of Theorem~\ref{th:th1} and is detailed in Supplementary Material~C4.

In the more general case of imperfect QRFs, where different frame orientations are assigned non-orthogonal states, both theorems fail, as detailed in Supplementary Material~D.

\section{Bell tests in QRFs} \label{sec:protocols}

\new{Non-local quantum states violating Bell's inequalities are a key quantum information resource~\cite{Brunner_bell_2014}, including applications in quantum cryptography~\cite{PhysRevLett.67.661,743501,PhysRevLett.95.010503,PhysRevLett.98.230501,Pironio_2009}, randomness generation~\cite{Pironio_2010,colbeck2011quantumrelativisticprotocolssecure,Colbeck_2011,arnon-Friedman2018}, and communication~\cite{PhysRevA.56.1201,PhysRevLett.89.197901,PhysRevLett.108.100401,ho2022}}. As discussed above, a change of QRF can lead to a decrease in entanglement and hence in the degree of non-locality of quantum states, at the expense of a corresponding increase in coherence. However, since non-locality is independent of single-particle coherence, \new{it may appear lost or gained under QRF changes}. If this hypothesis is indeed correct, it would suggest that those QRFs in which states manifest non-locality are \enquote{preferred}, which would challenge the view that all QRFs are equally legitimate perspectives.

In the following we will show that the aforementioned hypothesis is incorrect. A set of probabilities that violates Bell's inequalities in one QRF will violate them in any other QRF, since probabilities are invariant under QRF changes. While the resource for this violation may be entanglement of states in one QRF, it can be the non-separability of operations in another QRF. In this sense the QRF change can transfer the \enquote{non-locality of states} in one QRF into the \enquote{non-locality of operations} in another QRF.

Bell's local causality is a condition on joint probabilities
\begin{equation}
 p(a,b|x,y) = \sum_\lambda p(\lambda) \, p(a|x,\lambda) \, p(b|y,\lambda), 
\end{equation}
where $p(\lambda) \geq 0, \sum_\lambda p(\lambda) =1$, is a probability distribution of the ``hidden variable'' and $a$ and $b$ are outcomes in Alice's and Bob's laboratories when they choose settings $x$ and $y$, respectively. The probabilities that satisfy Bell's local causality satisfy Bell's inequalities. 

In quantum mechanics, Bell's inequalities can be violated, which demonstrates the incompatibility of Bell's local causality and quantum mechanics. To demonstrate the violation, Alice and Bob may share a pair of particles in an entangled quantum state and perform local operations on them. We refer to this as the \enquote{non-locality of states}. 

We now proceed to an analysis of how the resources required for a violation of Bell's inequalities change under QRF transformations. We begin our analysis in QRF $C$. Suppose that in this frame the quantum state $ \rho^{(C)}_{AB}$ of the composite system $A$ and $B$ is non-local in the sense that there are local POVM elements $ A^{(C)}_{a|x}$, $ B^{(C)}_{b|y}$ such that
\begin{equation}
 \label{nonlocal}
 \text{Tr}\left[ \rho^{(C)}_{AB} \left ( A^{(C)}_{a|x} \otimes B^{(C)}_{b|y} \right )\right] \neq \sum_\lambda p(\lambda) \, p(a|x,\lambda) \, p(b|y,\lambda).
\end{equation}

The change of QRF results in a refactorization of the Hilbert space from $\mathcal{H}_{AB}$ to $\mathcal{H}_{CB}$, encompassing both the states and the observables. In general, the initial entangled state in QRF $C$ can be mapped to a separable one in QRF $A$ while products of POVMs are in general mapped into non-separable POVMs 
\begin{align}
 S^{(C)\to(A)} \, {\rho}^{(C)}_{AB} \, (S^{(C) \to (A)})^{\dagger} & = {\rho}^{(A)}_C \otimes {\rho}^{(A)}_B \\
 (S^{(C) \to (A)})^{\dagger}\left ( {A}^{(C)}_{a|x} \otimes {B}^{(C)}_{b|y} \right )S^{(C)\to(A)} & = {D}^{(A)}_{a|x,b|y},
\end{align}
where, in general, ${D}^{(A)}_{a|x,b|y} \neq \sum_j q_j {C}^{(A)}_{j,a|x}\otimes {B}^{(A)}_{j,b|y}$, with $q_j \geq 0$ and $\sum_j q_j =1$. 

We note that as the change of QRF is a unitary transformation, the probabilities for measurement outcomes remain the same before and after the change. Hence, although the state in QRF $A$ is a local one, the probabilities still display non-locality since 
\begin{align}
 \text{Tr}\left[\left(\rho^{(A)}_{C} \otimes \rho^{(A)}_{B} \right) D^{(A)}_{a|x, b|y}\right] \neq \sum_\lambda p(\lambda) \,p(a|x,\lambda)\, p(b|y,\lambda). 
\end{align}
This is because the operations in QRF $A$ are in general not in a separable form. In fact, if a state in QRF $C$ violates a Bell's inequality for some local POVMs, then the whole set of transformed POVMs in QRF $A$ cannot be separable, because otherwise one would not violate Bell's inequalities in that frame. 
We refer to this violation as resulting from \enquote{non-locality of operations}. We conclude that the change from one to another QRF results in a change from \enquote{non-locality of states} to \enquote{non-locality of operations}.

To illustrate our analysis with an example consider the test of Clauser-Horne-Shimony-Holt (CHSH) inequality \cite{Clauser1969,Brunner_bell_2014}:
\begin{equation}
 \sum_{x,y=0}^1 p(a \oplus b =xy|x,y) \leq 3,
\end{equation}
where $p(a \oplus b =xy|x,y)$ is the probability to obtain outcomes $a \oplus b$ to be equal to the product $xy$, if the choice of the settings is the pair $x,y$. It is known that if Alice and Bob share an entangled state of qubits $A$ and $B$, as described by a reference qubit $C$,
\begin{equation}
 \ket{\psi}_{AB}^{(C)} = \frac{\ket{0}_A \ket{0}_B + \ket{1}_A\ket{1}_B}{\sqrt{2}},
 \label{eq:Bell_state}
\end{equation}
and measure eigenprojectors of a product of spins ${\sigma}_{\Vec{m}} \otimes {\sigma}_{\Vec{n}}$ along directions $\Vec{m}$ and $\Vec{n}$, they can maximally violate the CHSH inequality by $2+\sqrt{2}$.

Let us now analyse the CHSH test from the perspective of QRF $A$. Note that performing a QRF change to the frame of $A$, the state of $B$ and $C$ becomes
\begin{equation}
 \ket{\psi}_{BC}^{(A)} = \ket{0}_B \frac{\ket{0}_C + \ket{1}_C}{\sqrt{2}}.
\end{equation}
As we can expect from our results, the entanglement in the perspective of $C$ has been transformed into coherence in the perspective of $A$, and now Alice and Bob share a separable state. However, the observables also change. As an example, the observable $\sigma_x \otimes \frac{\sigma_x + \sigma_z}{\sqrt{2}} $, that is factorized in the perspective of $C$, here becomes non-separable, i.e.~$(S^{(C) \to (A)})^{\dagger} \, \sigma_x \otimes \frac{\sigma_x + \sigma_z}{\sqrt{2}} \, S^{(C)\to(A)} = 1/\sqrt{2} \, (\sigma_x \otimes \mathbb{1} - \sigma_y \otimes \sigma_y)$, in agreement with the general analysis given above.

\section{Conclusion} \label{sec:conclusions}

Quantum coherence and entanglement are fundamentally intertwined. In this paper, we explored this connection in the framework of QRFs, focusing on bipartite pure states. We proved the existence of a QRF invariant and showed that it can be decomposed in a way that captures the connection between coherence and entanglement. This was explicitly demonstrated for two pairs of quantifiers, showing that the sum of entanglement and subsystem coherence is conserved under QRF changes, implying that an increase of coherence under QRF change must come at the expense of a decrease in entanglement, and vice versa. Furthermore, we showed that a weaker trade-off holds for any possible pair of measures when there exists a QRF in which either the coherence or the entanglement vanishes.

Finally, we discussed violations of Bell's inequalities in different QRFs and highlighted the role of different quantum resources for the violation in different QRFs. As an example, we show that the entangled states and separable observables that violate the CHSH inequality are mapped to factorized states and non-separable observables that violate the same inequality.

In the Supplementary Material, we prove all statements and theorems from the main text and investigate imperfect QRFs.

\new{Future work could extend the present results to multipartite entanglement involving any number of parties}. Moreover, generalizing our results beyond the case of locally compact groups to the group of diffeomorphisms holds significant potential for insights into scenarios involving quantum gravity. Our findings suggest that while entanglement and subsystem coherence individually remain invariant under a classical diffeomorphism, their combined sum is preserved under \emph{quantum} diffeomorphisms. This extension would imply that $\mathcal{C}+\mathcal{E}$ is an observable within the framework of general relativity, opening up exciting avenues for exploring the implications of this insight in the realm of quantum gravity.

\begin{acknowledgments}
CC, AA, and ACdlH thank the organizers and participants of the \enquote{Sejny Summer Institute 2022} where this project started. ACdlH thanks Stefan Ludescher and Markus Müller for helpful discussions on imperfect QRFs.
PC is supported by the National Science Centre (NCN, Poland) within the Preludium Bis project (No. 2021/43/O/ST2/02679) and the National Agency for Academic Exchange (NAWA, Poland) as a part of the Preludium Bis NAWA program (No. BPN/PRE/2023/1/00001).
This research was funded in whole or in part by the Austrian Science Fund (FWF) [10.55776/F71] and [10.55776/COE1]. For open access purposes, the author has applied a CC BY public copyright license to any author accepted manuscript version arising from this submission. Funded by the European Union - NextGenerationEU. This publication was made possible through the financial support of the ID 62312 grant from the John Templeton Foundation, as part of The Quantum Information Structure of Spacetime (QISS) Project (qiss.fr). The opinions expressed in this publication are those of the authors and do not necessarily reflect the views of the John Templeton Foundation.
\end{acknowledgments}

\bibliography{ref}

\appendix
\onecolumngrid

\section{QRF transformations are incoherent operations} \label{app: incoherent operations}
In this Supplemental Material we show that QRF transformations, defined in Eq.~\eqref{eq:QRFT}, are incoherent operations. We focus on the case of finite Hilbert spaces, to avoid possible divergences in the coherence \new{quantifiers}. Note, however, that the result also holds in the infinite dimensional case. Given a Hilbert space and an orthonormal basis $\lbrace |i\rangle \rbrace_{i=0,\cdots,d-1}$, a state is called incoherent in that basis if it can be written as $\sigma=\sum_i \sigma_i |i \rangle \langle i|$ (see Ref.~\cite{Baumgratz2014} for an overview). We refer to the set of incoherent states as $\mathcal{I}$. In turn, a CPTP map $\Lambda$ is called incoherent if, given its Kraus decomposition, i.e.~$\Lambda[\rho] = \sum_i K_i \rho K_i^\dagger$, each Kraus operator maps incoherent states to incoherent states, i.e.~$\forall i \ K_i \mathcal{I} K_i^\dagger \subset \mathcal{I}$. As a simpler case, a unitary transformation is called incoherent if it maps incoherent states to incoherent states. We now show that this is the case for QRF transformations.

Let us consider a generic incoherent state in the perspective of $C$, i.e.
\begin{equation}
 \rho_{AB}^{(C)} = \sum_{g,g^\prime} P_{g,g^\prime} \ket{g}\bra{g} \otimes \ket{g^\prime}\bra{g^\prime}.
\end{equation}
The QRF transformation of Eq.~\eqref{eq:QRFT} can be written in the discrete case as $S = \sum_g \ket{g^{-1}}\bra{g} \otimes U(g^{-1})$. By direct application, and remembering that $U(g) \ket{h}=\ket{g\circ h}$, we find
\begin{align}
 S \rho_{AB}^{C} S^\dagger &= \sum_{g,g^\prime} P_{g,g^\prime}\ket{g^{ -1}}\bra{g^{-1}} \otimes \ket{g^{-1} \circ g^\prime} \bra{g^{-1} \circ g^\prime} \\
 &=\sum_{g,g^\prime} \ P_{g^{-1},g^{-1} \circ g^\prime}\ket{g}\bra{g} \otimes \ket{g^\prime} \bra{g^\prime},
\end{align}
where the second line has been obtained after two changes of variables. It is then clear that the resulting state is still incoherent, since the effect of the QRF transformation has only been the permutation of the diagonal elements.

This means that QRF transformations are indeed incoherent. By unitarity, they also have to preserve the coherence of the bipartite state~\cite{Peng2016}. This characterization provides a deep insight into the structure of the studied transformations. Within the QRF formalism, no additional coherence of the bipartite state can be created - \textit{the total coherence is preserved} for any coherence measure. Furthermore, new possibilities for studying quantum resources associated with QRF transformations arise by using results from resource theory. One example could be the application of Strelstov's theorem~\cite{Strelstov2015}, which concerns a wide family of coherence and entanglement measures based on distances, and allows one to upper bound the amount of entanglement present in the system by its coherence.

Note, however, that the coherence of each subsystem can change through a QRF transformation, and this is what is studied in the main text.

\section{Conservation of diagonal elements up to permutation} \label{app:Lemma}
\subsection{Continuous case}
We present here a lemma that leads to the main results of our work (Theorem~1 and Theorem~2). Given a group $G$, a QRF transformation $S$ associated to the group, an initial and final state $\ket{\psi}_{AB}^{(C)}$, $\ket{\psi}_{BC}^{(A)} = S\ket{\psi}_{AB}^{(C)}$, and the respective reduced density matrices $\rho_A^{(C)} = \Tr_B[\ket{\psi}_{AB}^{(C)} \bra{\psi}]$, $\rho_C^{(A)} = \Tr_B[\ket{\psi}_{BC}^{(A)} \bra{\psi}]$, we show that the diagonal elements of the two reduced density matrices are the same up to permutation. We can express this statement as a set equality:
\begin{equation}
 \left\{\braket{g|\rho_A^{(C)}|g} \Big | \, g \in G\right\} = \left\{\braket{g|\rho_C^{(A)}|g} \Big | \, g \in G\right\}.
\label{eq:lemma}
\end{equation}
This can be proven as follows. Let us start with the state in the perspective of $C$ defined in Eq.~\eqref{eq:state_perspective_C}. Tracing out subsystem $B$ we get
\begin{equation}
 \rho_A^{(C)}= \int dg \, \int \, dg^\prime \, \int dh \,\psi^{(C)}_{AB}(g,h)\psi^{*(C)}_{AB}(g^\prime,h)\, \ket{g}\bra{g^\prime}.
\end{equation}
The set of its diagonal elements is thus given by
\begin{equation}
 D^{(C)} = \left\{\braket{g|\rho_A^{(C)}|g} =\int \, dh \,|\psi^{(C)}_{AB}(g,h)|^2\ \ \Big | \, g \in G\right\}.
\end{equation}
Similarly, for $\rho_C^{(A)}$ we obtain
\begin{equation}
 D^{(A)} = \left\{\braket{g|\rho_C^{(A)}|g} =\int \, dh \,|\psi^{(A)}_{BC}(g,h)|^2\ \ \Big | \, g \in G\right\}.
\end{equation}
From the definition of a QRF transformation in Eq.~(6) we have
\begin{equation}
 \braket{g|\rho_C^{(A)}|g} = \int \, dh \,|\psi^{(A)}_{BC}(g,h)|^2 = \int \, dh \,|\psi^{(C)}_{AB}(g^{-1},g^{-1} \circ h) |^2 = \int \, dh \,|\psi^{(C)}_{AB}(g^{-1},h) |^2 = \braket{g^{-1}|\rho_A^{(C)}|g^{-1}}.
\end{equation}
Since for each element of the group $G$ there exists a unique inverse in $G$, then $D^{(C)} = D^{(A)}$, and Eq.~\eqref{eq:lemma} is proven. 

The above observation can be used to derive the conservation laws in the main text, and the derivation is discussed in detail in Supplemental Material~\ref{app:proofTh}. Note that this result holds in the case of ideal reference frames, but there are also more general types of quantum reference frames for which this result does not hold, as discussed in Supplemental Material~\ref{app: imperfect quantum reference frames}.

\subsection{Discrete case}

\label{sec:discrete_proof}
We provide here an alternative proof for discrete groups, which highlights the role of permutations in QRF transformations. Let us again consider three parties, $A$, $B$ and $C$, and the most general discrete pure state with respect to $C$:
\begin{equation}
 \ket{\psi}_{AB}^{(C)} = \sum_{g,h} C_{g,h} \ket{g}_A \ket{h}_B.
\end{equation}
Here, $C_{g,h}$ is such that the state is normalized, namely $ \sum_g (CC^\dagger)_{g,g} = 1$.
Applying the QRF transformation, we get the state with respect to $A$,
\begin{equation}
 \ket{\psi}_{BC}^{(A)} = \sum_{g,h} C_{g,h} \ket{g^{-1}}_C \ket{g^{-1}\circ h}_B = \sum_{g,h} C_{g^{-1}, g^{-1}\circ h} \ket{g}_C \ket{h}_B := \sum_{g,h} D_{g,h} \ket{g}_C \ket{h}_B.
\end{equation}
Tracing out the second subsystem from the perspective of $C$, we get
\begin{equation}
 \rho_A = Tr_B[\ket{\psi}_{AB}^{(C)}\bra{\psi}] = \sum_{g,h} (C C^\dagger)_{g,h} \ket{g}\bra{h},
\end{equation}
and the diagonal part $\rho_{A,d} = \sum_g (C C^\dagger)_{g,g} \ket{g} \bra{g}$.
Analogously, starting from the state with respect to $A$, performing a QRF change to $C$, and applying a partial trace, we get
\begin{equation}
 \rho_{C,d} = \sum_g (DD^\dagger)_{g,g} \ket{g}\bra{g}.
\end{equation}
Here, we want to show that
\begin{equation}
 (DD^\dagger)_{g,g} = (CC^\dagger)_{\pi(g) \pi(g)},\ \forall g,
\end{equation}
where $\pi(\cdot)$ is a permutation of the elements of $G$. Note that, by the definition of QRF transformations, we have
\begin{equation}
 D_{g,h} = C_{g^{-1},{g^{-1}} \circ h}.
\end{equation}
A QRF change thus consists of applying a permutation $P^{(i)}$ on the columns of the density matrix, such that each element $g$ gets mapped to its inverse $g^{-1}$, and a controlled permutation on the rows, such that each row undergoes a different permutation (this is the effect of the controlled unitary on the second system). That is,
\begin{equation}
 D = P^{(i)} \left(\sum_{j=1}^{|G|} A^{(j)} C P^{(j)}\right),
\end{equation}
where $P^{(i)}$ and $P^{(j)}$ are permutations matrices, $A^{(j)} = \text{diag}(\{\delta_{jk}\}_{k=1,\dots,|G|})$ are the control elements, i.e.~diagonal matrices with only one element equal to $1$ and the other elements being zero. The permutation matrices are orthogonal, i.e.~$P P^T = 1$, and the control matrices are symmetric, that is, $A = A^T$. Overall, this yields
\begin{equation}
 DD^\dagger = P^{(i)} \left(\sum_{jl} A^{(j)} C P^{(j)} P^{(l)T} C^\dagger A^{(l)}\right)P^{(i)T}.
\end{equation}
Note that given matrices $M$ and $N$, if $N = P M P^T$, then $N_{g,g} = M_{\pi(g), \pi(g)}$, that is, the permutation acts in such a way that the diagonal elements are permuted between each other. Hence we obtain
\begin{equation}
 (DD^\dagger)_{g,g} = \left(\sum_{j,l} A^{(j)} C P^{(j)} P^{(l)T} C^\dagger A^{(l)}\right)_{\pi(g), \pi(g)},
\end{equation}
where the permutation $P^{(i)}$ is the one that sends each element to its inverse, namely, $\pi(g)$ is the inverse of $g$.

Furthermore, we use the property that when $A^{(j)}$ and $A^{(l)}$ are applied to the left and right of a matrix, they select one element, namely $A^{(j)} M A^{(l)} = M_{j,l} S^{(j,l)}$, where $S^{(j,l)}$ is the matrix that has a 1 in position $j,l$, and zero otherwise: $S^{(j,l)}_{a,b} = \delta_{j,a} \delta_{l,b}$. As a result,
\begin{equation}
 (DD^\dagger)_{gg} = \sum_{j,l} (CP^{(j)}P^{(l)T}C^\dagger)_{j,l} S^{(jl)}_{\pi(g),\pi(g)} = (CC^\dagger)_{\pi(g),\pi(g)},
\end{equation}
which proves our statement.

\section{Proofs of theorems}\label{app:proofTh}
\subsection{Proof of Theorem 1: Entanglement entropy and coherence entropy} \label{app:proofEE}
To formulate the first conservation relation we use the \textit{entanglement entropy} and the \textit{relative entropy of coherence}. Given a pure bipartite state $\ket{\psi}_{AB}$ and a density matrix $\rho$ of arbitrary dimension, the measures are defined as~\cite{NielsenChuang, Baumgratz2014}
\begin{align}
 \mathcal{E}_e\left[|\psi\rangle_{AB}\right]&=\mathcal S\left[\rho_A\right], \label{eq:ent_entropy} \\
 \mathcal{C}_e\left[\rho \right]&= \mathcal S\left[\rho_d\right]-\mathcal S\left[\rho \right],
\end{align}
where $\mathcal S[\rho] = -\Tr\left[\rho \log \rho\right]$ is the von Neumann entropy, $\rho_A = \Tr_B \left [\ket{\psi}_{AB}\bra{\psi}\right ]$ is the reduced density matrix of subsystem $A$, and the subscript $d$ denotes the diagonal part of the state in the group basis, that is,
$\braket{g|\rho_{d}|g^\prime} = \delta(g^{-1}\circ g^\prime) \braket{g|\rho|g}$.

If we calculate the coherence of the reduced density matrix $\rho_A^{(C)} = \Tr_B \left[\ket{\psi}_{AB}^{(C)}\bra{\psi}\right]$, and of $\rho_C^{(A)}$, we find
\begin{align}
 \mathcal{C}_e\left[\rho_{A}^{(C)}\right] + \mathcal{E}_e\left[\ket{\psi}_{AB}^{(C)}\right] &= \mathcal S\left[\rho_{A,d}^{(C)}\right], \\
 \mathcal{C}_e\left[\rho_{C}^{(A)}\right] + \mathcal{E}_e\left[\ket{\psi}_{BC}^{(A)}\right] &= \mathcal S\left[\rho_{C,d}^{(A)}\right].
\end{align}
We now want to show that these two quantities are equal, as in Eq.~\eqref{eq:conservation1} in the main text. This happens when $\mathcal S\left[\rho_{A,d}^{(C)}\right] = \mathcal S\left[\rho_{C,d}^{(A)}\right]$, which can be proved as follows. By construction, $\rho_{A,d}^{(C)}$ has the same diagonal elements as $\rho_A^{(C)}$, while all the other elements are zero, hence the diagonal elements of $\rho_A^{(C)}$ are the eigenvalues of $\rho_{C,d}^{(A)}$ (and analogously for $\rho_{C,d}^{(A)}$ and $\rho_C^{(A)}$). Furthermore, the lemma of Eq.~\eqref{eq:lemma} states that $\rho_A^{(C)}$ and $\rho_C^{(A)}$ have the same diagonal elements up to a permutation, and hence $\rho_{A,d}^{(C)}$ and $\rho_{C,d}^{(A)}$ have the same eigenvalues up to a permutation. Since the von Neumann entropy of a state depends only on the set of eigenvalues of the state, it follows that the two von Neumann entropies are the same, proving Eq.~\eqref{eq:conservation1} in the main text.

\subsection{Proof of Theorem 1: Linear entropy and $l^2$ coherence} \label{app:proofLE}
As for the second conservation relation, we consider as coherence and entanglement \new{quantifiers} the $l^2$-norm of coherence $\mathcal{C}_{l^2}$ and the linear entropy of the subsystem $\mathcal{E}_l$, respectively. \new{Note that the latter can be treated as a rescaled square of the entanglement concurrence.} Once again, we focus here on discrete groups and finite Hilbert spaces. These \new{quantifiers} are given by~\cite{Angelo2004, Baumgratz2014}
\begin{align}
\mathcal{E}_l \left [|\psi\rangle_{AB}\right ]&=1-\Tr\left[\rho_A^2\right], \\
 \mathcal{C}_{l^2}\!\left [\rho\right ]&=\sum_{i\neq j} |\rho_{ij}|^2.
\end{align}
The above \new{$l^2$ coherence} can be easily decomposed into two terms: $\mathcal{C}_{l^2}\!\left [\rho \right ]=\sum_{i,j}|\rho_{ij}|^2-\sum_{i}|\rho_{ii}|^2$. The first term is actually the state's purity $\Tr \left[\rho^2 \right]$ and the latter is sometimes referred to as the \enquote{classical purity}. By the lemma of Eq.~\eqref{eq:lemma}, the second quantity is conserved under QRF transformations. In other words, the quantity
\begin{equation}
\mathcal{C}_{l^2}\!\left [\rho^{(C)}_A \right]-\Tr \left [ \left(\rho_A^{(C)}\right)^2 \right]=-\sum_{g \in G} |(\rho^{(C)}_A)_{g,g}|^2
\end{equation}
is invariant under QRF transformations.
Finally, adding $1$ to both sides of the above equation and writing the same equation in the reference frame of $A$, one finds
\begin{equation}
\mathcal{C}_{l^2}\!\left [\rho^{(C)}_A \right] + \mathcal{E}_l \left [|\psi\rangle_{AB}^{(C)}\right ] = \mathcal{C}_{l^2}\!\left [\rho^{(A)}_C \right] + \mathcal{E}_l \left [|\psi\rangle_{BC}^{(A)}\right ],
\end{equation}
which is Eq.~\eqref{eq:conservation2}
in the main text. 

We have thus found two conservation relations, using two different pairs of measures for coherence and entanglement. Note, however, that it is not possible to find the same conservation law for any possible pair of measures: an example is given in the next section, where $\mathcal{C}_{l^2} + \mathcal{E}_e$ is not conserved, and not even the weaker condition $\Delta \mathcal{C}_{l^2} \Delta \mathcal{E}_{e}\leq 0$ holds.

Nevertheless, the lemma of Eq.~\eqref{eq:lemma} can be used to prove other entanglement-coherence trade-offs, possibly in a more convoluted form. To give an example, it is possible to find another conservation relation for two qubits. Let us consider the geometric measure of entanglement~\cite{Guhne2009} $\mathcal{E}_g \left[\ket{\psi_{AB}}\right]=1-\frac{1}{2}\left(1 + \sqrt{1-4 \det(\rho_A)}\right)$ and the geometric coherence~\cite{Strelstov2015} $\mathcal{C}_g \left [\rho \right]=1/2 \left(1-\sqrt{1-4|\rho_{01}|^2}\right)$. We can write
\begin{align}
\mathcal{E}_g \left [|\psi\rangle^{(C)}_{AB} \right ]&=1-\frac{1}{2}\left (1+\sqrt{1+2 \, \mathcal{C}_g \! \left [\rho^{(C)}_A \right ] \left (1-2\, \mathcal{C}_g \left [\rho^{(C)}_A \! \right ] \right )-4\rho_{00}\rho_{11}} \right ) \\
 &=1-\frac{1}{2}\left (1+\sqrt{1+2 \, \mathcal{C}_g \! \left [\rho^{(C)}_A \right ] \left (1-2\, \mathcal{C}_g \left [\rho^{(C)}_A \! \right ] \right )- I} \right ),
\end{align}
where $I$ is a QRF invariant, due to the lemma of Eq.~\eqref{eq:lemma}. Solving the above equation for $I$, one could arrive at yet another entanglement-coherence trade-off relation.

\subsection{Uniqueness proof}
\new{
We prove here that once an entanglement (or coherence) \new{quantifier} is chosen, there cannot be more than one coherence (or entanglement) \new{quantifier} such that their sum is QRF invariant. We focus on the proof with a fixed entanglement \new{quantifier}, but a completely analogous proof holds for the case with a fixed coherence \new{quantifier}.

Specifically, we prove that given $\mathcal{E}_A$ and $\mathcal{C}_A$ such that their sum is invariant under QRF changes, i.e. 
\begin{equation}
    \mathcal{E}_A + \mathcal{C}_A = k_A,
\end{equation}
where $k_A$ is a QRF invariant, then there cannot exist another $\mathcal{C}_B \neq \mathcal{C}_A$, such that $\mathcal{E}_A + \mathcal{C}_B = k_B$, with $k_B$ a QRF invariant.

Let us prove this by contradiction: assuming that $\mathcal{C}_B$ exists such that $\mathcal{E}_A + \mathcal{C}_B = k_B$ entails that 
\begin{equation}
    \mathcal{C}_B = \mathcal{C}_A + K,
\end{equation} 
where $K = k_B - k_A$ is a QRF invariant.

From this relation, it follows that $\mathcal{C}_B$ is a coherence \new{quantifier} if and only if $K$ is. But $K$ is by assumption a QRF invariant, and since coherence changes under QRF transformations, $K$ cannot be a coherence \new{quantifier} (for example, it can be non-zero for incoherent states). As a result, there cannot exist $\mathcal{C}_B \neq \mathcal{C}_A$ such that $\mathcal{E}_A + \mathcal{C}_B = k_B$, and this concludes the proof.
}
\subsection{Counterexample for a generic inequality}\label{app:counterexample}
We have shown that there are two conserved quantities under QRF transformations, namely Eqs.~\eqref{eq:conservation1} and \eqref{eq:conservation2} in the main text. They imply that the following two inequalities hold: $\Delta \mathcal{C}_e \Delta \mathcal{E}_e\leq 0$ and $\Delta \mathcal{C}_{l^2} \Delta \mathcal{E}_{l}\leq0$. This, however, does not mean that for any possible pair of measures, one can obtain an analogous inequality. For example, it does not hold true that for any state $\Delta C_{l^2} \Delta \mathcal{E}_e\leq 0$. Here we provide a counterexample of a state and a QRF transformation for which the inequality is violated.

Let us consider the cyclic group $G=\mathbb{Z}_3$, a basis associated to the group elements $\{\ket{0},\ket{1},\ket{2}\}$, and a unitary representation of the group $U(g)$, acting as $U(g) \ket{g^\prime} = \ket{(g+g^\prime)\mod 3}$.
A counter example is an initial state in the perspective of $C$, i.e.~$\ket{\psi}_{AB}^{(C)} = \sum_{i,j=0}^3 \psi^{(C)}_{i,j} \ket{ij}_{AB}$, where the coefficients are (approximately) given by
\begin{equation*}
\psi^{(C)} = \begin{bmatrix}
+0.20 - 0.05i & -0.22 + 0.30i & -0.28 + 0.12i \\
-0.36 + 0.21i & -0.15 - 0.24i & +0.25 + 0.12i \\
-0.41 + 0.31i & +0.04 - 0.09i & -0.08 - 0.34i \\
\end{bmatrix}.
\end{equation*}
After the QRF change, one finds the state
\begin{equation*}
\psi^{(A)} = \begin{bmatrix}
+0.20 - 0.05i & -0.22 + 0.30i & -0.28 + 0.12i \\
+0.25 + 0.12i & -0.36 + 0.21i & -0.15 - 0.24i \\
+0.04 - 0.09i & -0.08 - 0.34i & -0.41 + 0.31i \\
\end{bmatrix}.
\end{equation*}
By explicit computation of the measures, one finds $\mathcal{E}^{(C)}_e=0.76$, $\mathcal{C}^{(C)}_{l^2}=0.23$, $\mathcal{E}^{(A)}_e=0.69$, $\mathcal{C}^{(A)}_{l^2}=0.17$, hence $\Delta \mathcal{E}_e\Delta \mathcal{C}_{l^2} = 0.004 >0$.

\subsection{Proof of Theorem 2}\label{app:proofth2}
We provide here a proof of Theorem~2, 
which states that, given a bipartite pure state $\ket{\psi}_{AB}^{(C)}$ in the frame of $C$ and its transformed version $\ket{\psi}_{BC}^{(A)} = S^{(C)\to (A)}\ket{\psi}_{AB}^{(C)}$ in the frame of $A$, for any possible choice of coherence and entanglement measures $\mathcal{C}$ and $\mathcal{E}$, if there exists a frame in which $\mathcal{C} \left [\rho_A^{(C)} \right ] = 0$ or $\mathcal{E} \left [\ket{\psi}_{AB}^{(C)} \right ] =0 $, with $\rho_A^{(C)} = \Tr_B \left[\ket{\psi}_{AB}^{(C)}\bra{\psi}\right]$ and similarly for $\rho_C^{(A)}$, then
\begin{equation}
 \left (\mathcal{C}\left [\rho_C^{(A)} \right ] - \mathcal{C} \left [\rho_A^{(C)} \right ] \right )\left (\mathcal{E} \left [\ket{\psi}_{BC}^{(A)} \right ] - \mathcal{E} \left [\ket{\psi}_{AB}^{(C)} \right ] \right )\leq 0.
\end{equation}
 From now on, we will adopt the same convention of the main text, and call $\mathcal{C}^{(C)} = \mathcal{C} \left [\rho_A^{(C)} \right ]$, $\mathcal{E}^{(C)}=\mathcal{E} \left [\ket{\psi}_{AB}^{(C)} \right ]$ for short, and equivalently for $A$. 

Let us start from the $\mathcal{C}^{(C)} = 0$ case. Since for any coherence measure it holds $\mathcal{C}^{(A)}\geq 0$, then we have to prove that \new{$\mathcal{E}^{(C)} \geq \mathcal{E}^{(A)}$. We will do so by proving that $\ket{\psi}_{AB}^{(C)}$ can be converted to $\ket{\psi}_{BC}^{(A)}$ by a LOCC transformation}. Let us consider the reduced density matrix $\rho_A^{(C)}$. Since $\mathcal{C}^{(C)}=0$, it is diagonal and hence its vector of diagonal elements $\Vec{d}^{(C)}$ coincides with its vector of eigenvalues $\Vec{\lambda}^{(C)}$, which are the Schmidt coefficients of $\ket{\psi}_{AB}^{(C)}$. In an equation: $\Vec{d}^{(C)} = \Vec{\lambda}^{(C)}$. Moreover, by the Schur-Horn Theorem~\cite{MatrixAnalysis}, for any Hermitian matrix, and in particular for $\rho_C^{(A)}$, it holds $\Vec{d}^{(A)} \preceq \Vec{\lambda}^{(A)}$, where $\preceq$ stands for weak majorization. Moreover, due to the lemma of Eq.~\eqref{eq:lemma}, we know that $\Vec{d}^{(C)} = \Vec{d}^{(A)}$, hence it must be that $\Vec{\lambda}^{(C)} \preceq \Vec{\lambda}^{(A)}$, where these vectors are the Schmidt coefficients of $\ket{\psi}_{AB}^{(C)}$ and $\ket{\psi}_{BC}^{(A)}$, respectively. 

\new{Finally, by Nielsen's theorem~\cite{Nielsen1999}, since the Schmidt coefficients of $\ket{\psi}_{BC}^{(A)}$ majorize the ones of $\ket{\psi}_{AB}^{(C)}$, this means that $\ket{\psi}_{AB}^{(C)}$ can be transformed into $\ket{\psi}_{BC}^{(A)}$ by a LOCC, and hence $\mathcal{E}^{(C)} \geq \mathcal{E}^{(A)}$. This concludes the first part of the proof.}

Let us investigate now the case where $\mathcal{E}^{(C)}=0$. We want to show that coherence can only decrease, i.e.~that $\mathcal{C}^{(A)}\leq \mathcal{C}^{(C)}$. To prove this, we show that the map between the initial subsystem and the final one is incoherent, when the initial state is factorized.
Given the reduced density matrices in the two perspectives, namely $\rho_A^{(C)}$ and $\rho_C^{(A)}$, these two states are related through a CPTP map $\phi \left[\rho_A^{(C)}\right] = \rho_C^{(A)}$ that corresponds to the QRF change. Explicitly the map is given by
\begin{equation}
 \phi\left[\rho_A^{(C)}\right] = \text{Tr}_B \left[S\, \rho_A^{(C)} \otimes \ket{\psi}^{(C)}_B \bra{\psi} \, S^\dagger\right],
\end{equation}
since the initial state is factorized by assumption, and the form of $S$ is given in Eq.~(6). This map can be decomposed in terms of Kraus operators $E(g)$, satisfying $\phi\left[\rho\right] = \int dg \, E(g) \rho E(g)^\dagger$, where one can choose $E(g)= \braket{g|S|\psi}_B^{(C)}$. 

Such an operation is incoherent iff, given any diagonal state $\rho_D = \int dh \, \rho_h \ket{h}\bra{h}$, it remains diagonal after the action of each Kraus operator, i.e.~$E(g) \rho_D E(g)^\dagger$ is diagonal $\forall g$. This can indeed be proven explicitly: From the expression of the Kraus operators and the definition of the QRF transformation, one can find
\begin{equation}
 E(g) = \int dg^\prime \braket{g \circ g^\prime | \psi}_B^{(C)} \, \ket{g^{\prime -1}}\bra{g^\prime}
\end{equation}
and replacing it in the incoherent operation condition, one gets
\begin{equation}
 E(g) \rho_D E(g)^\dagger = \int dh \, \rho_h \braket{g\circ h | \psi}_B\braket{\psi|g\circ h} \, \ket{h^{-1}} \bra{h^{-1}},
\end{equation}
which is clearly diagonal, as the proof requires. Hence, since $\phi$ is an incoherent operation, it cannot increase coherence, and it must be that $\mathcal{C}^{(A)}\leq \mathcal{C}^{(C)}$. As a consequence, it holds $\Delta \mathcal{C} \Delta \mathcal{E} \leq 0$, when $\mathcal{E}^{(A)}=0.$

\section{Imperfect quantum reference frames} \label{app: imperfect quantum reference frames}

In the main text, we have restricted our analysis to $L^2(G)$ quantum reference frames, i.e.~QRFs that carry the regular representation of the symmetry group. These are referred to in the literature as perfect~\cite{delaHamette2020quantumreference} or ideal~\cite{delahamette2021perspectiveneutral} quantum reference frames. In the following, we show that Theorem~1 that we derived for ideal frames no longer holds for non-ideal QRFs, i.e.~$\mathcal{C}_e + \mathcal{E}_e$ and $\mathcal{C}_{l^2} + \mathcal{E}_l$ are not conserved under an imperfect QRF change. This happens because the type of imperfect QRFs we discuss here is characterized by an overcomplete basis of group elements, and consequently the QRF transformation between any two perspectives is more involved.

To see why, let us first give a brief introduction to the perspective-neutral approach to QRFs~\cite{vantrietvelde_switching_2018, vanrietvelde2018change, hoehn2019trinity, Krumm_2021, hoehn2021internal, delahamette2021perspectiveneutral, hoehn2023quantum}, which represents an alternative formalism for internal QRFs. Consider a system of interest $S$ and a QRF $R$, both transforming under a given symmetry group $G$. The group is taken to be an arbitrary unimodular Lie group, that is, the left- and right-invariant Haar measures of the Lie group coincide. Given a kinematical description of both systems, one can assign a quantum state to $S$ and $R$ in the \emph{kinematical} Hilbert space, which takes the product form $\ch_\kin=\ch_R \otimes \ch_S$. $\ch_\kin$ is assumed to carry a unitary tensor product representation of $G$, i.e.~for all $g \in G$ and $\ket{\psi_\kin}$, we have $U_{RS}(g) \ket{\psi_\kin} = U_R(g) \otimes U_S(g) \ket{\psi_\kin}$, where $g \mapsto U_R(g)$ and $g \mapsto U_S(g)$ are strongly continuous unitary representations. 
We shall require the existence of a \emph{system of coherent states} $\{ U_R, \ket{\phi(g)}_R \}$ such that $G$ acts transitively on the coherent state system. That is, for all $g,g'\in G$, we have $\ket{\phi(g'g)}_R= U_R(g')\ket{\phi(g)}_R$, and, in particular, $\ket{\phi(g)}_R=U_R(g)\ket{\phi(e)}_R$. The coherent states $\ket{\phi(g)}_R$ are interpreted as the different possible orientation states of the $G$-frame, where $g$ is called the \emph{quantum frame orientation}. In order for the frame orientation states to admit a consistent probabilistic interpretation, the representation $U_R(g)$ and the seed coherent state $\ket{\phi(e)}$ are required to give rise to a \emph{resolution of the identity} 
\begin{align}
 \int_G dg\,\ket{\phi(g)}\!\bra{\phi(g)}_R=n\,\mathbf{1}_R,
\end{align}
for $n>0$ a constant.

An \emph{ideal reference frame} is characterized by carrying the regular \emph{representation} of $G$, in which case $\ch_R\simeq L^2(G,dg)$. In this case, $G$ acts regularly on itself, and the frame orientation states are orthogonal and thus perfectly distinguishable, i.e.~$\braket{g|g'}=\delta(g^{-1} \circ g')$. A reference frame that does not carry the regular representation is referred to as \emph{non-ideal.}

The physical Hilbert space contains all \emph{physical} states, that is, all states invariant under the gauge group action $U_{RS}$. They are of the form $U_{RS}(g)\,\ket{\psi_{\rm phys}}=\ket{\psi_{\rm phys}}\,,\forall\,g\in G$.
They can be constructed by applying the \emph{coherent group averaging \enquote{projector}} $\Pi_{\rm phys}:=\int_G dg\,U_{RS}(g)$.
Then, with a slight abuse of notation (see Ref.~\cite{delahamette2021perspectiveneutral} for a detailed discussion), we have $\ket{\psi_{\rm phys}}:=\Pi_{\rm phys}\,\ket{\psi_{\rm kin}}$.

In order to reduce from a physical \enquote{perspective-neutral} state to an internal QRF perspective, one can apply the \emph{Schrödinger reduction map} $\mathcal{R}_{\mathbf{S},R}(g): \mathcal{H}_{\rm phys} \to \mathcal{H}_{S,g}^{\rm phys}$, which maps states from the physical Hilbert space to the description of the system $S$ relative to the frame $R$ being in orientation $g\in G$:
\begin{align}
\mathcal{R}_{\mathbf{S},R}(g) = \langle\phi(g)|_R \otimes \mathbf{1}_S. \label{eq:reduction_map}
\end{align}
The image of this map is the \emph{physical system Hilbert space} $\mathcal{H}_{S,g}^{\rm phys}$, which contains all relative states of $R$ that result from conditioning a gauge-invariant, physical state on the frame $R$ being in orientation $g$. Thus, we can define \emph{relative physical states} as $|\psi_S^{\rm phys}(g)\rangle := \mathcal{R}_{\mathbf{S},R}(g)\,|\psi_{\rm phys}\rangle$. The latter satisfy the covariance property $U_S(g')|\psi_S^{\rm phys}(g)\rangle = |\psi_S^{\rm phys}(g'g)\rangle$ and allow to rewrite physical states as
\begin{align}
|\psi_{\rm phys}\rangle=\int_G dg'|\phi(g')\rangle_R\otimes|\psi_S^{\rm phys}(g')\rangle. \label{eq:physical_state}
\end{align}

Finally, let us see how to change perspective from one frame to another. Consider two frame systems $R_1$ and $R_2$, and the system of interest $S$. The kinematical Hilbert space is given by $\mathcal{H}_{\rm kin} = \mathcal{H}_{R_1} \otimes\mathcal{H}_{R_2} \otimes \mathcal{H}_S$ and the physical space is the image of the kinematical space under the coherent group averaging projector. To reduce to the perspective of the internal frame $R_i$ being in orientation $g_i, i=1,2$, we apply the Schrödinger reduction map $\mathcal{R}_{\mathbf{S},R_i}(g_i)$. Since this map is invertible when its domain is restricted to $\mathcal{H}_{\rm phys}$, it is now possible to change perspectives by passing via the perspective-neutral structure. The QRF change that maps the physical states of $S$ relative to $R_1$ being in orientation $g_1$ to the states of $S$ relative to $R_2$ being in orientation $g_2$ is given by 
\begin{align}
 V_{R_1\to R_2}(g_1,g_2):=\mathcal{R}_{\mathbf{S},R_2}(g_2)\cdot\mathcal{R}_{\mathbf{S},R_1}^{-1}(g_1), \label{eq:imperfectframechange}
\end{align}
mapping from $\mathcal{H}_{R_2S,g_1}^{\rm phys}$ to $\mathcal{H}_{R_1S,g_2}^{\rm phys}$.\\

Having reviewed the perspective-neutral approach to QRFs, let us, for simplicity, choose a concrete example for the symmetry group $G$, namely the smallest finite non-Abelian group $S_3$. Take three systems $A$, $B$, and $C$, such that each has a Hilbert space $\mathbb{C}^3$ and carries the representation that maps $S_3$ onto the triangle. That is, $S_3$ acts as permutations on the three basis vectors $|i\rangle,\ i=0,\dots,2$, which form an orthonormal basis of $\mathbb{C}^3$. The coherent state system can be constructed from a general seed state $|\phi(e)\rangle =\alpha |0\rangle +\beta|1\rangle +\gamma |2\rangle$, where $\alpha,\beta,\gamma\in\mathbb{C}$. The conditions on the seed state are summarized as follows:
\begin{align}\label{eq:seedstatecond}
 &\alpha, \beta, \gamma \text{ are pairwise different,} \nonumber\\
 &|\alpha|^2+|\beta|^2+|\gamma|^2=1, \\
 &\alpha\beta^*+\alpha^*\beta^*+\alpha\gamma^*+\alpha^*\gamma^*+\gamma\beta^*+\gamma^*\beta^*=0.\nonumber
\end{align}
The physical Hilbert space for $N=3$ systems for $G=S_3$ is five-dimensional. We can reduce into the perspectives of systems $A$ and $C$ by applying the respective Schrödinger reduction map $\mathcal{R}_{\mathbf{S},A}(e) = \langle\phi(e)|_A \otimes \mathbf{1}_{BC}$ or $\mathcal{R}_{\mathbf{S},C}(e) = \langle\phi(e)|_C \otimes \mathbf{1}_{AB}$.\\

Let us now give a concrete example in the case of imperfect frames for $S_3$ in which the conservation of $\mathcal{C}_e + \mathcal{E}_e$ and $\mathcal{C}_{l^2} + \mathcal{E}_l$ under changes of frame does not hold. Take the kinematical state of $A$, $B$, and $C$ in $\ch_\kin=(\mathbb{C}^3)^{\otimes 3}$ to be 
\begin{align}
 \ket{\psi_\kin}_{CAB}=\ket{0}_C \otimes \frac{1}{\sqrt{3}} (i\ket{0}_A+\ket{1}_A-\ket{2}_A) \otimes \ket{2}_B.
\end{align}
Applying the group averaging projector $\Pi_\phys=\frac{1}{\sqrt{|S_3|}}\sum_{g\in S_3} U_{CAB}(g)$, we get the corresponding physical state
\begin{align}
 \ket{\psi_\phys}_{CAB}=\frac{1}{\sqrt{6}}\frac{1}{\sqrt{3}}\big(
 &i\ket{002}+\ket{012}-\ket{022} 
 +i\ket{110}+\ket{120}-\ket{100} 
 +i\ket{221}+\ket{201}-\ket{211} \nonumber\\
 +&i\ket{001}+\ket{021}-\ket{011} 
 +i\ket{220}+\ket{210}-\ket{200} 
 +i\ket{112}+\ket{102}-\ket{122}\big)_{CAB}. 
\end{align}
Let us now reduce to the perspectives of systems $C$ and $A$, respectively. We choose the following seed state:
\begin{align}
 \ket{\phi(e)}=\frac{1}{\sqrt{2}}(\ket{0}+i\ket{1}).
\end{align}
It is easy to check that this state satisfies conditions \eqref{eq:seedstatecond}.
We apply the Schrödinger reduction map
\begin{align}
 \mathcal{R}_{\mathbf{S},C}(e)=\sqrt{3}\frac{1}{\sqrt{2}}(\bra{0}-i\bra{1})_C\otimes \mathbb{1}_{AB}
\end{align}
to the physical state. Note that we added here a multiplicative factor of $\sqrt{3}$ in order to render the relative physical states normalized. We obtain the relative physical state of $A$ and $B$, given $C$ in orientation $e$:
\begin{align}
 \mathcal{R}_{\mathbf{S},C}(e) \ket{\psi_\phys}_{CAB}=\frac{\sqrt{3}}{6}(i\ket{00}+i\ket{01}+\ket{10}-\ket{11}+2\ket{12}-i\ket{20} +\ket{21}+(i-1)\ket{22})_{AB}.
\end{align}
Similarly, we obtain the relative physical state of $B$ and $C$, given $A$ in orientation $e$:
\begin{align}
 \mathcal{R}_{\mathbf{S},A}(e) \ket{\psi_\phys}_{CAB}=\frac{\sqrt{3}}{6}(2i\ket{01}+(-1-i)\ket{20}+(1+i)\ket{21}+2\ket{12})_{CB}.
\end{align}
We are now in the position to compare the diagonal elements of the reduced density matrices:
\begin{align}
 &\rho_A^{(C)}=\Tr_B[\rho_{AB}^{(C)}]=\frac{1}{12}(2\ketbra{0}{0}+6\ketbra{1}{1}+4\ketbra{2}{2})\text{ + off-diagonals,}\\
 &\rho_C^{(A)}=\Tr_B[\rho_{CB}^{(A)}]=\frac{1}{12}(4\ketbra{0}{0}+4\ketbra{1}{1}+4\ketbra{2}{2})\text{ + off-diagonals.}
\end{align}

We can then see that after an imperfect QRF change the diagonal elements of $\rho_A^{(C)}$ and $\rho_C^{(A)}$ are not equal up to a permutation. This shows that the lemma of Eq.~\eqref{eq:lemma} does not hold anymore for imperfect QRFs. As a consequence, one can check by direct computation that the conservation laws of Theorem~1 do not hold. In fact:
\begin{align}
 \Delta \mathcal{C}_e +\Delta \mathcal{E}_e &= S \left [\rho_{A,d}^{(C)} \right ]- S \left [ \rho_{C,d}^{(A)} \right ] = \left(\frac{\log 2}{2}+\frac{\log 3}{3}+\frac{\log 6}{6}\right)-\log 3 \approx -0.09 \neq 0, \\
 \Delta \mathcal{C}_{l^2} +\Delta \mathcal{E}_{l} &= \sum_i \left (\left |\left (\rho_C^{(A)}\right )_{i,i} \right |^2 - \left |\left (\rho_A^{(C)}\right )_{i,i} \right |^2 \right ) = -\frac{1}{18} \neq 0.
\end{align}

Let us now briefly consider the inequalities $\Delta \mathcal{C}_e \Delta \mathcal{E}_e <0$ and $\Delta \mathcal{C}_{l^2} \Delta \mathcal{E}_l <0$. While one can show that these inequalities are satisfied for ideal QRFs, as is done in the main text, this is generally not the case for imperfect QRFs that carry a coherent state system. Let us provide a counterexample here.

The general form of a reduced physical state is given by 
\begin{align}
\ket{\psi}^{(A)}_{BC}=
&a\cdot(\alpha^\ast\ket{00}_{BC} + \beta^*\ket{11}+\gamma^\ast\ket{22}) \nonumber \\ +(1/\sqrt{2}) [ 
&b\cdot (\alpha^\ast (\ket{11}+\ket{22})+\beta^\ast(\ket{00}+\ket{22}) + \gamma^\ast (\ket{00}+\ket{11})) \nonumber \\
+ &c\cdot (\alpha^\ast (\ket{10}+\ket{20})+\beta^\ast(\ket{00}+\ket{21}) + \gamma^\ast (\ket{11}+\ket{12})) \nonumber \\
+ &d\cdot (\alpha^\ast (\ket{01}+\ket{02})+\beta^\ast(\ket{10}+\ket{12}) + \gamma^\ast (\ket{20}+\ket{21})) \nonumber \\
+ &e\cdot (\alpha^\ast (\ket{12}+\ket{21})+\beta^\ast(\ket{02}+\ket{20}) + \gamma^\ast (\ket{01}+\ket{10}))].
\end{align}
One counterexample for Theorem~1 and the above-mentioned inequalities is generated by the following parameters:
\begin{align}
 &(\alpha,\beta,\gamma)=\frac{1}{\sqrt{3}}(i,1,-1) \text{ and } \\
 &(a,b,c,d,e)=\left(\frac{39}{86} - \frac{29i}{69}, -\frac{77}{171} + \frac{37i}{85}, \frac{39}{89} + \frac{i}{43}, \frac{39}{272} - \frac{8i}{95}, -\frac{3}{41} - \frac{i}{513}\right),
\end{align}
for which
\begin{align}
 \Delta \mathcal{C}_e +\Delta \mathcal{E}_e=-0.138 \text{ and } \Delta \mathcal{C}_e \Delta \mathcal{E}_e = 0.005, \\
 \Delta \mathcal{C}_{l^2} +\Delta \mathcal{E}_l=-0.453 \text{ and } \Delta \mathcal{C}_{l^2} \Delta \mathcal{E}_l = 0.005.
\end{align}

Regarding Theorem~2, let us briefly explain why the assumptions do not hold in the case of imperfect QRFs that carry a coherent state system. Note that the change of QRF in Eq.~\eqref{eq:imperfectframechange} is a map from one physical system Hilbert space to another. Any state relative to one QRF is obtained from applying $\Pi_{\rm phys}$ to a kinematical state, which strongly entangles the state, and conditioning on a frame state $\ket{\phi(g)}$. Using Eqs.~\eqref{eq:reduction_map} and \eqref{eq:physical_state}, this gives rise to states of the form
\begin{align}
 \ket{\psi}_{BC}^{(A)} \equiv\mathcal{R}_{\mathbf{S},A}(g) \circ\Pi_{\rm phys} \ket{\psi_{\rm kin}}= \int dg' \braket{\phi(g)|\phi(g')} \ket{\psi^{\rm phys}_{BC}(g')}
\end{align}
for states relative to $A$, and analogously for $C$.
For any imperfect frame with overcomplete basis $
\{ \ket{\phi(g)} \}_{g\in G}$, that is, $\braket{\phi(g)|\phi(g')} \neq \delta(g'^{-1} \circ g)$, the relative state will still contain a certain degree of entanglement. We can thus not find separable states in $\mathcal{H}^{\rm phys}_{BC,g}$ or $\mathcal{H}^{\rm phys}_{AB,g}$. For the same reason, one can also not find any relative states with zero coherence. Thus, the statement of Theorem~2 does not apply to the case of such imperfect QRFs. 

\end{document}